\documentclass{article}
\usepackage{amsmath,amsfonts}
\usepackage{algorithmic}
\usepackage{algorithm}
\usepackage{array}
\usepackage[caption=false,font=normalsize,labelfont=sf,textfont=sf]{subfig}
\usepackage{textcomp}
\usepackage{stfloats}
\usepackage{url}
\usepackage{verbatim}
\usepackage{graphicx}
\usepackage{cite}
\usepackage{hyperref}
\usepackage{cleveref}
\usepackage{amsmath, amssymb, amsthm} 
\usepackage{graphicx} % Required for inserting images

\title{Sparse Matrix Coding for URLLC}
\author{Yifei Yang}
\date{May 2024}

\begin{document}

\maketitle

\section{Background of coding for URLLC}
\subsection{Introduction}
URLLC (Ultra-Reliable Low-Latency Communications) is a fundamental concept within the realm of 5G and Internet of Things (IoT) technologies, specifically designed to meet the stringent requirements of applications with demanding latency and reliability needs, which \(10^{-5}\) block error rate (BLER) within \(1ms\) period.

The communication requirements of URLLC encompass several critical aspects, the first being ultra-high reliability.  URLLC mandates exceptional reliability in the communication link to ensure error-free data transmission during critical moments.  Applications such as remote surgeries and intelligent traffic control rely on uninterrupted communication to avoid severe consequences. Another pivotal requirement of URLLC is extremely low latency.  Real-time applications like autonomous vehicles and remote industrial control demand prompt responses, making the minimization of communication delay imperative.  URLLC ensures swift data transmission from sender to receiver, facilitating timely decision-making or action.

Moreover, URLLC must support massive connectivity due to the proliferation of Internet of Things (IoT) devices.  Efficient resource allocation and management are necessary to accommodate numerous devices concurrently, ensuring that each device receives adequate bandwidth and reliability. Additionally, URLLC requires optimized scheduling and synchronization for applications that involve multiple devices working collaboratively, such as distributed energy management and smart cities.  This optimization maximizes overall performance by ensuring that data transmission occurs accurately and timely.

To fulfill these communication requirements, URLLC leverages various advanced technologies and strategies.  Mini-slot technology reduces air interface latency by optimizing time resource utilization, enhancing data transmission efficiency.  Optimization of the Modulation and Coding Scheme (MCS) and Channel Quality Indicator (CQI) tables ensures reliable data transmission under low-latency conditions. Furthermore, URLLC integrates edge computing, optimal path collaboration, user plane acceleration, and end-to-end latency monitoring within its ultra-low latency solution architecture.  Edge computing minimizes data transmission delay by processing tasks at the network edge, while optimal path collaboration selects the best network path for efficient data transmission. Acceleration of the user plane optimizes data transmission procedures and end-to-end latency monitoring enables real-time identification and resolution of issues. URLLC also introduces the Transmission Time Interval (TTI) structure to further reduce latency and enhance reliability. This structure provides flexible management of data transmission time intervals, improving efficiency and reliability. Meanwhile, URLLC supports flexible frame structures to cater to diverse application scenarios and requirements. This flexibility allows for adjustment of data transmission rates and latencies based on specific needs, ensuring compliance with URLLC's stringent requirements.

However, existing technologies, despite their efforts to meet URLLC's business needs, inevitably increase the complexity of communication architecture and coding. In order to meet the QoS requirements of URLLC more precisely, in addition to optimizing the communication structure, it is also necessary to carry out in-depth improvement and innovation on the coding level of short packet information. This requires not only a deeper understanding of existing technologies, but also the continuous exploration and development of new coding strategies and methods to ensure efficient and reliable communication.

\subsection{Related Works}

Shim et al.\cite{SVC2018twc} propose an innovative signal processing approach to solve the problem of efficient transmission of short packet information in wireless communication. They suggested a sparse representation of short packet information in sparse vectors, capturing essential characteristics, and removing redundancy. This compact representation enhances transmission efficiency. To further compress sparse vectors, they introduce supercomplete dictionaries, representing vectors as a linear combination of a few atoms. This reduces dimensionality and enhances compression, improving anti-interference ability. Then, sparse vector coding (SVC) is placed on the orthogonal frequency division multiplexing (OFDM) grid for a spread spectrum effect, carefully designed for accurate decoding. At the receiving end, compressed sensing algorithms recover the original information without increasing complexity.

SVC exhibits excellent low payload on pilot \cite{pilotless2019LWC} and anti-interference ability through spread spectrum communication and compressed sensing. Experimentally, SVC meets URLLC requirements at lower SNR than traditional coding schemes. They provide an error formula, offering a theoretical basis for evaluating the performance of SVC.

Sato and colleagues \cite{SATO2021JIOT,SATO22022JCC,Yang2023LCOMM} have further improved and upgraded SVC encoding. They augmented the existing indices in SVC with high-order modulation symbols, allowing for the transmission of more information without increasing the complexity of information decoding and they call this scheme as SSC (Sparse Superimposed Coding). However, this direct augmentation method is prone to inter-symbol leakage and crosstalk. Consequently, in subsequent papers, they implemented additional rotation processing for each symbol requiring mapping. This ensures that the Euclidean distance between each symbol remains maximal, thereby enhancing interference resistance. Experimental results indicate that compared to traditional SVC, this approach leads to an approximately 2dB increase in the required SNR for transmission with the same amount of information.

In another endeavor, Author in \cite{optimalMat2023LWC} sought to enhance encoding by refining the coding dictionary to meet the demands of short packet communication. They devised two greedy algorithms to minimize column correlations within the coding dictionary, ultimately obtaining the optimal coding dictionary through optimization problem solving. Simulation results demonstrate that, compared to existing approaches, SVC's block error rate (BLER) performance can be significantly enhanced.

\section{Sparse Matrix Coding}
\subsection{SMC Encoding}

Sparse Matrix Coding (SMC) is an efficient data representation method that aims to restructure multiple sparse vectors into high-dimensional indices. These new indices are then used as sparse values in a high-dimensional matrix. The primary objective of this approach is to represent and store data in a more efficient and compact way.

Using the example of two user data packets, we can leverage the mapping relationship of Sparse Vector Coding (SVC) to transform them into sparse vectors\cite{SVC2018twc,SATO2021JIOT,SATO22022JCC}. Let's assume that the sparse indices for User 1 and User 2 are (4,7) and (3,6) respectively. This indicates that the nonzero elements in User 1's sparse vector are located at the 4th and 7th positions, while those in User 2's sparse vector are located at the 3rd and 6th positions.

By combining these indices into matrix indices (4,3) and (7,6), we construct a high-dimensional matrix where the positions at the 4th row and 3rd column, as well as the 7th row and 6th column, are assigned non-zero values. This approach results in a sparse matrix in which most elements are zero, with only non-zero values at specific indices. This matrix effectively encapsulates the information from both users. If we collapse this matrix along the column direction, we obtain User 1's sparse vector, and collapsing along the row direction gives us User 2's sparse vector. This representation not only preserves the original  information, but also significantly reduces the required storage space by leveraging sparsity.

In SVC, a super-complete dictionary is typically used to compress high-dimensional vectors. Similarly, we can apply a similar compression technique to this sparse matrix. Specifically, we can compress the matrix by multiplying it with a complete dictionary of both front and back. This compression reduces the data's dimension and complexity, leading to improved processing efficiency.

\begin{figure}[!t]
\centering
\includegraphics[width=3in]{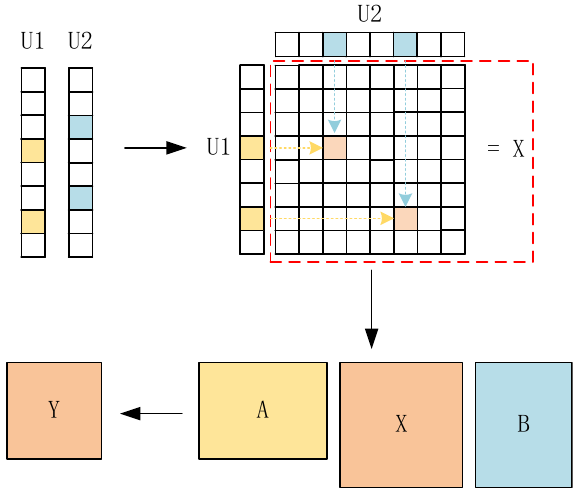}
\centering
\caption{SMC Encoding Procedures}
\label{encode}
\end{figure}

\begin{figure}[!t]
\centering
\includegraphics[width=3in]{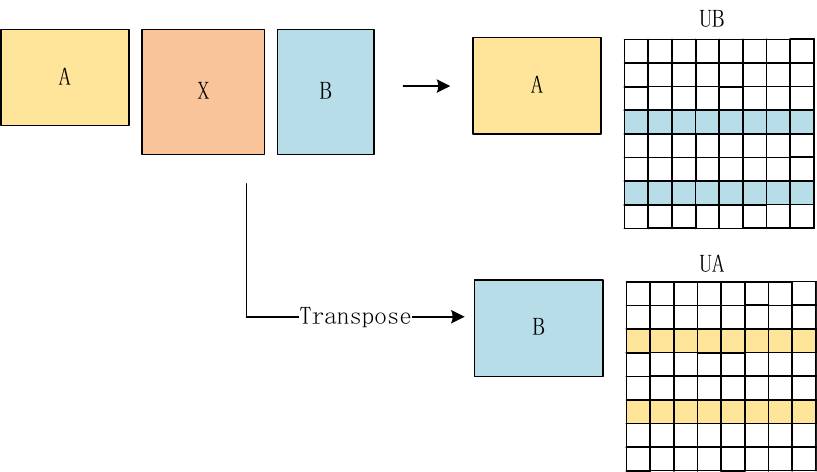}
\centering
\caption{SMC Decoding Procedures}
\label{decode}
\end{figure}

%Overall, SMC, through the combination of SVC mapping and sparse matrix representation, offers an effective method for handling and storing high-dimensional data. This approach not only reduces the dimensionality and complexity of the data but also enhances data representation and storage efficiency of the data, providing a powerful tool for processing large-scale datasets.

\subsection{SMC Decoding}

Compared to the SVC decoding process, the core of SMC lies in the handling of sparse matrices. Given that the spreading dictionary \(\mathbf{A}\) and the time-spreading dictionary \(\mathbf{B}\) are known at the receiver, decoding can be performed using this prior knowledge. Taking the example of two users, if we directly apply the spreading dictionary \(\mathbf{A}\) to recover the received signal \(\mathbf{Y}\), \(\mathbf{Y}\) can be considered as the product of \(\mathbf{A}\) and \(\mathbf{UB}\). Since U itself is a sparse matrix, \(\mathbf{UB}\) will be a row-sparse matrix, with the row indices of its sparse blocks corresponding to the SVC sparse value indices of U1.

Furthermore, if \(\mathbf{Y}\) is transposed and vectorized, this is equivalent to transposing and vectorizing \(\mathbf{UB}\). Subsequently, by expanding \(\mathbf{A}\) to \(\mathbf{\Phi}\) (as shown in \eqref{eq:vec_encode_eq}), we can obtain a block-sparse vector \(\mathbf{ub}\) and its corresponding dictionary An. At this point, the problem transforms into a block-sparse signal recovery problem, which can be solved using existing block-sparse recovery algorithms. Finally, combining \(\mathbf{ub}\) in the row direction, we can obtain the matrix \(\mathbf{UB}\).

As \(\mathbf{A}\) and \(\mathbf{B}\) are random Bernoulli matrices, we compute the column sums of the absolute values in \(\mathbf{UB}\) to determine the indices of the information transmitted by User U1. Notably, during the decoding process for user U1, we can also simultaneously decode the information for user U2. This can be achieved by calculating the inner product of each sparse row in \(\mathbf{UB}\) with each row in \(\mathbf{B}\), where the index corresponding to the maximum inner product value is the decoding result for user U2.

Additionally, since the dictionaries are known, the dimension of \(\mathbf{X}\) is also known, which means that the length of the sparse blocks in \(\mathbf{ub}\), obtained through vectorization, is also known. This information is crucial for block-sparse signal recovery and can effectively reduce the computational complexity of the recovery algorithm.

For user U2, decoding can be performed directly, in addition to using user U1. This is because \(\mathbf{Y}\) can be represented equivalently as the product of \(\mathbf{UA}\) and \(\mathbf{B}\). The sparsity of U results in \(\mathbf{UA}\) being a column-sparse matrix. By transposing \(\mathbf{Y}\), we can obtain the product of \(\mathbf{B}^H\) and \(\mathbf{UA}^H\), which is similar to the decoding problem for user U1. Therefore, the same decoding steps can be applied to obtain the information for user U2. Similarly, by calculating the inner product of each sparse column in \(\mathbf{UA}\) with each column in \(\mathbf{A}\), we can obtain the information for user U1. This mutual decoding and verification mechanism not only enhances the flexibility of decoding but also allows parallel decoding, avoiding queuing delays, and ensuring efficient information transmission.

If we merge this matrix along the column direction, we can obtain the sparse vector for user 1, while merging along the row direction gives the sparse vector for user 2. This representation not only preserves the information of the original data, but also significantly reduces the required storage space by leveraging sparsity.

In SVC, a super-complete dictionary is typically used to compress a high-dimensional vector. Analogously, we can perform a similar compression operation on this sparse matrix. Specifically, we can achieve matrix compression by multiplying a complete dictionary on both sides of the matrix. The benefit of this approach is that it further reduces the dimension and complexity of the data, thus improving processing efficiency.

In general, SMC provides an effective means of handling and storing high-dimensional data by combining the SVC mapping and the representation of sparse matrices. By reducing dimension and complexity, this method improves the efficiency of data representation and storage, offering a robust solution for processing large-scale datasets. 

Recovering block-sparse signals typically requires further expansion and computation of two extension dictionaries, leading to a significant increase in spatial and computational complexity. To mitigate this complexity, we propose setting the time-spreading dictionary \(\mathbf{B}\) equal to the transpose of the spreading dictionary \(\mathbf{A}\). This approach ensures that the dictionaries used for block-sparse recovery by U1 and U2 are identical, thus simplifying the overall computation and reducing complexity.

\section{Performance of Sparse Matrix Coding}
% \subsection{Code Efficiency of SMC}

\subsection{Reliability of SMC under Rayleigh Channel}

We can reformulate the encoding process into a matrix-based representation as shown in \eqref{eq:mat_encode_eq}. 
\begin{align}
\mathbf{Y}=\mathbf{H}\mathbf{A}\mathbf{X}\mathbf{A}^H+\mathbf{N} \label{eq:mat_encode_eq},
\end{align}
where \(\mathbf{X}_{n\times n}\) represents the encoded sparse matrix, \(\mathbf{A}_{m\times n}\) is the dictionary matrix, \(\mathbf{H}_{m\times m}\) denotes the complex Gaussian channel matrix, which is \(diag(h_1,\cdots,h_m)\) and every \(h_i\) follows complex Gaussian distribution, \(\mathbf{Y}_{m\times m}\) is the received matrix, and \(\mathbf{N}_{m\times m}\) represents the noise matrix within the channel.

Due to the fact that \(\mathbf{X}\) is a sparse matrix, indicating that the majority of its elements are zeros, multiplication between \(\mathbf{X}\) and \(\mathbf{A}^H\) inevitably results in a row-sparse matrix, where most rows consist mainly of zeros. In this context, \(\mathbf{HA}\) can be used to recover \(\mathbf{X}\mathbf{A}^H\). Typically, the received signal is vectorized, which converts it from a one-dimensional array to a vector. Additionally, the dictionary matrix is tensor-expanded, increasing its dimensionality. This procedure is equivalent to transposing \(\mathbf{X}\mathbf{A}^H\) and then column-vectorizing it, converting the matrix into a long column vector. After vectorization of \eqref{eq:mat_encode_eq}, the resulting formula is as follows:
\begin{align}
    Vec(\mathbf{Y}^H)=((\mathbf{HA})\otimes \mathbf{I_m})\cdot Vec(\mathbf{A}\mathbf{X}^H)+Vec(\mathbf{N}^H).
    \label{eq:vec_encode_eq}
    %\\\mathbf{y} &= \mathbf{\Phi}\cdot \mathbf{x}+\mathbf{n} 
\end{align}

We can also rewrite \eqref{eq:vec_encode_eq} as \(\mathbf{y} = \mathbf{\Phi}\cdot \mathbf{x}+\mathbf{n}\) where 
\(\mathbf{y}_{mm\times 1}=Vec(\mathbf{Y}^H),~\mathbf{\Phi}_{mm\times nm}=((\mathbf{HA})\otimes \mathbf{I_m}),~\mathbf{x}_{nm\times 1}=Vec(\mathbf{A}\mathbf{X}^H),~\mathbf{n}=Vec(\mathbf{N}^H)\).
During this process, sparse signals on the sparse vector exhibit a block pattern, referred to as block-sparse signals. This occurs because, after column-vectorization, the non-zero elements that were originally distributed by rows now coalesce into blocks, forming a block sparse structure.

It is worth noting that while \(\mathbf{\Phi}\) represents the tensor-expanded version of \(\mathbf{A}\), the fundamental information of the column remains unchanged. In our problem, we consider the originally row-sparse matrix \(\mathbf{X}\mathbf{A}^H\), which is an \(n\times m\) matrix. After transposition and column-vectorization, it transforms into a block-sparse vector with a dimension of \(nm\times 1\). This vector is then segmented into n independent blocks based on the \(m\) elements. Similarly, the expanded dictionary has \(nm\) columns, corresponding to each element of the block-sparse vector. Consequently, the dictionary is also divided into \(n\) blocks, with each block containing \(m\) columns as shown in \eqref{eq:Phi_divide}.
\begin{align}
\mathbf{\Phi}&=
\begin{bmatrix}
  \mathbf{\Phi}_1&  \mathbf{\Phi}_2&  \cdots &\mathbf{\Phi}_n \notag
\end{bmatrix}
\\&=
\begin{bmatrix}
  h_1a_{11}\mathbf{I}_m&  h_1a_{12}\mathbf{I}_m& \cdots  &  h_1a_{1n}\mathbf{I}_m\\
  h_2a_{21}\mathbf{I}_m&  h_2a_{22}\mathbf{I}_m& \cdots  &  h_2a_{2n}\mathbf{I}_m \\
  \vdots &  \vdots&  &  \vdots  \\
  h_ma_{m1}\mathbf{I}_m&  h_ma_{m2}\mathbf{I}_m& \cdots &  h_ma_{mn}\mathbf{I}_m 
\end{bmatrix},\label{eq:Phi_divide}
\end{align}
where \(\mathbf{\Phi}_i\) represents sub-matrix blocks of the dictionary partitioned according to \(n\) columns, \(a_{ij}\) represents the element at the \(i\)-th row and \(j\)-th column in \(\mathbf{A}\) and \(h_i\) is the channel gain on \(i\)-th path in \(\mathbf{H}\).

Since the block-sparse nature of the signals we aim to solve for, most of the signal values are close to zero, while the non-zero values are concentrated within discrete blocks composed of rows from \(\mathbf{X}\mathbf{A}^H\). In the context of row-sparse matrices, each row contains either all zeros or sparse values exclusively. Consequently, after transposition and column-vectorization, the boundaries of each sparse block in the resulting block-sparse signal are inherently known. This block-sparsity pattern is commonly encountered in practical applications, such as texture recognition in image processing and multi-path signal analysis in wireless communications. Leveraging this unique structure, we can simplify the problem by only recovering the indices of the non-zero blocks, rather than precisely reconstructing the value of each signal sample.

In our context, rather than recovering the entire signal, we focus on identifying the block indices. Given the block-sparse nature of the signal and the known block boundaries, we segment the dictionary \(\mathbf{A}\) accordingly and approximate \(\mathbf{Y}\) using linear combinations of these blocks. This approach not only reduces computational complexity, but also improves the efficiency of the solution. Consequently, a block-greedy algorithm can be employed to address this problem. This algorithm involves matching \(\mathbf{Y}\) against the segmented dictionary, searching for the most relevant block at each iteration and incorporating its index and corresponding block into the sparse vector recovery solution. This approach enables successful recovery with simplicity and effectiveness, making it a highly viable solution.

Assuming the maximum column correlation in the dictionary \(\mathbf{A}\) is \(\mu\), the correlation between each block in the expanded dictionary \(\mathbf{\Phi}\) can also be represented using matrix inner products. Therefore, we can get \eqref{eq:beta_mat} where we use \(\beta_{\mathbf{\Phi_i,\Phi_j}}\) to represent the matrix inner products between \(i\)-th block and \(j\)-th block in \(\mathbf{\Phi}\), 
\begin{align}
\beta_{\mathbf{\Phi_i,\Phi_j}} &=<{\mathbf{\Phi_i} },{\mathbf{\Phi_j}}>\notag
\\&= \sum_{k=1}^{m} <\frac{\mathbf{\phi}^{(k)}_i }{||\mathbf{\phi}^{(k)}_i||_2},\frac{\mathbf{\phi}_j^{(k)} }{||\mathbf{\phi}_j^{(k)}||_2} >\notag
\\&=\left\{\begin{matrix} m, ~~~i=j \\m\cdot \mu,~~~i\ne j\end{matrix}\right.\label{eq:beta_mat},
\end{align}
where \(\mathbf{\phi}^{(k)}_i\) is the \(k\)-th column vector in \(i\)-th block of \(\mathbf{\Phi}\).

For the purpose of facilitating subsequent descriptions and with some abuse of notation, we define the inner product between matrices and vectors as
\begin{align*}
\left \langle {\mathbf{\Phi_i}} ,\mathbf{y}\right \rangle 
=\sum_{k=1}^{m} <\frac{\mathbf{\phi}^{(k)}_i }{||\mathbf{\phi}^{(k)}_i||_2},\frac{\mathbf{y}}{||\mathbf{y}||_2}  >.
\end{align*}

In fact, performing an inner product between the vectorized \(\mathbf{y}\) and the blocked division of \(\mathbf{\Phi}\) is equivalent to performing an inner product between each column of HA and the corresponding column of matrix \(\mathbf{Y}\). Therefore, when the matching block is inner-producted with \(\mathbf{y}\), the resulting value is
\begin{align*}
\left \langle {\mathbf{\Phi_i}} ,\mathbf{y}\right \rangle 
&=\sum_{k=1}^{m} <\frac{\mathbf{\phi}^{(k)}_i }{||\mathbf{\phi}^{(k)}_i||_2},\frac{\mathbf{y}}{||\mathbf{y}||_2}  >
\\&=\left\{\begin{matrix} m, ~~~matched \\m\cdot \mu,~~~others\end{matrix}\right. .
\end{align*}

As evident from the preceding discussion, block matching exhibits significant advantages in terms of correlation compared to conventional single-column matching. When performing block matching, there is always a gap proportional to the block length between matched and unmatched blocks. This implies that the difference in correlation between matched and unmatched blocks increases as the block length increases. This property makes block matching particularly suitable for information recovery and identification. When dealing with large amounts of information, being able to clearly distinguish between matched and unmatched blocks greatly enhances the efficiency of information recovery and identification.

Similar to SVC, we can get 
% \begin{lemma}
    the upper bound of BLER of SMC is 
    \begin{align*}
        1-\left(1-(\frac{(\beta_{\mathbf{B}^{(t)},\mathbf{B}^{(t)}}-\beta_{\mathbf{B}^{(t)},\mathbf{B}^{(i)}})^2\left(\sum_{k=1}^n s_k \right)^2}{2\sigma ^2})^{-m}-(\frac{(\beta_{\mathbf{B}^{(t)},\mathbf{B}^{(t)}})^2\left(\sum_{k=1}^n s_k \right)^2}{\sigma ^2}
)^{-m}\right)^{Kn}
    \end{align*}

% \end{lemma}

Due to the employment of a greedy algorithm, the information within sparse blocks of sparse vectors is iteratively recovered based on relevance. Therefore, the successful decoding formula can be defined as
\begin{align}
P_{success} &= P(\mathbf{B}^{(1)},\cdots ,\mathbf{B}^{(K)})\notag
\\&=\prod_{k=2}^KP(\mathbf{B}^{(k)}|\mathbf{B}^{(k-1)})P(\mathbf{B}^{(1)})
\approx \prod_{k=1}^K P(\mathbf{B}^{(k)})\label{eq:succ_prob}
\end{align}
where
\begin{align}P(\mathbf{B}^{(t)})=\int P(\mathbf{B}^{(t)}|\mathbf{h})P(\mathbf{h})d\mathbf{h}=\mathbb{E}_{\mathbf{h}} [P(\mathbf{B}^{(t)}|\mathbf{h})].
\label{eq:prob_block}
\end{align}

Therefore, the decoding probability given the knowledge of \(\mathbf{h}\) is
\begin{align}
P(\mathbf{B}^{(t)}|\mathbf{h})
&=P(|\left \langle {\Phi_{\mathbf{B}^{(t)}}} ,\mathbf{y}\right \rangle| > \max_{i\ne t}|\left \langle {\Phi_{\mathbf{B}^{(i)}}} ,\mathbf{y}\right \rangle|)\notag
\\&=\prod_{i\ne t}^{n} P(|\left \langle {\Phi_{\mathbf{B}^{(t)}}} ,\mathbf{y}\right \rangle| > |\left \langle {\Phi_{\mathbf{B}^{(i)}}} ,\mathbf{y}\right \rangle|)
\label{eq:p_b_h}
\end{align}
where we turn maximum into the form of consecutive multiplication of probabilities.

Now, we categorize and discuss the results of the inner product between matching and non-matching blocks. It is important to note that, due to the assumption that the columns of the original dictionary \(\mathbf{A}\) have been normalized, we can get \(||\phi||_2=||h||_2\) same as in \cite{SATO22022JCC}. Therefore, when the selected block is a matching block, both \(\mathbf{\Phi}\) and \(\mathbf{y}\) inherently contain \(\mathbf{h}\). Consequently, during the inner product operation, \(\mathbf{h}\) also participates in the inner product. Then, we can get the following equation
\begin{align}
|\left \langle {\Phi_{\mathbf{B}^{(t)}}} ,\mathbf{y}\right \rangle|
&=|\sum_{i=1}^{m} <\frac{{\phi}_k^{(t)} }{||{\phi}_k^{(t)}||_2},\frac{\mathbf{y}}{||\mathbf{y}||_2}    >|
\notag
\\&=
\left |   
||\mathbf{h}||_2^2  \sum_{k=1}^{n} s_k\cdot \beta_{{\mathbf{B}^{(t)}},{\mathbf{B}^{(t)}}}
+||\mathbf{h}||_2z_n
\right |\label{eq:inner_prod_matched}
\end{align}
and in mismatched case, we also can get 
\begin{align}
|\left \langle {\Phi_{\mathbf{B}^{(i)}}} ,\mathbf{y}\right \rangle|
&=|\sum_{i=1}^{m} <\frac{{\phi}_k^{(i)} }{||{\phi}_k^{(i)}||_2},\frac{\mathbf{y}}{||\mathbf{y}||_2}    >|
\\&=
\left |   
||\mathbf{h}||_2^2  \sum_{k=1}^{n} s_k\cdot \beta_{{\mathbf{B}^{(t)}},{\mathbf{B}^{(i)}}}
+||\mathbf{h}||_2z_p
\right |, \label{eq:inner_prod_mismathed}
\end{align}
where \(\sum_{k=1}^{n} s_k\) represents the sum of sparse values within each sparse column of the row-sparse matrix \(\mathbf{XA}^H\), \(\beta_{{\mathbf{B}^{(t)}},{\mathbf{B}^{(t)}}}\) denotes the block correlation for the matched block while \(\beta_{{\mathbf{B}^{(t)}},{\mathbf{B}^{(i)}}}\) denotes the block correlation for the mismatched, and \(z_n,z_p\) are the result of inner product between noise vector and selected block. Since the fact that all elements in the noise matrix are independently and identically distributed according to a Gaussian distribution, performing an inner product between its columns and the normalized columns does not alter the distributional characteristics. Under these circumstances, both \(z_n\) and \(z_p\) also follow a Gaussian distribution, which is 
\begin{align}
z_n,z_p&=\sum_{k=1}^{m} <\frac{{\phi}_k }{||{\phi}_k||_2},n_k  >\notag\\&\sim N(0,\frac{\sigma^2}{2})\label{eq:z_np_distribution}
\end{align}

Therefore, by substituting \eqref{eq:inner_prod_matched} and \eqref{eq:inner_prod_mismathed} into \eqref{eq:p_b_h} and further applying the absolute value inequality for scaling by \eqref{eq:scale_matched} and \eqref{eq:scale_mismatched}, we can obtain 
\begin{align}
&P\left(|\left \langle {\Phi_{\mathbf{B}^{(t)}}} ,\mathbf{y}\right \rangle|>|\left \langle {\Phi_{\mathbf{B}^{(i)}}} ,\mathbf{y}\right \rangle|\right)\notag
\\&\ge
 P\left\{\left(\left|\left\|\boldsymbol{h}\right\|_2\left|\sum_{k=1}^ns_k\beta_{{\mathbf{B}^{(t)}},{\mathbf{B}^{(t)}}}\right|-\left|z_n\right|\right|
>\left|z_p\right|\right.\right.
+\left\|\boldsymbol{h}\right\|_2\left.\left|\sum_{k=1}^ns_k\beta_{\mathbf{B}^{(t)},\mathbf{B}^{(i)}}\right|\right) \notag
\\&=
P\left\{ \left\| \boldsymbol{h} \right\|_2 \left( \left| \sum_{k=1}^n s_k \beta_{\mathbf{B}^{(t)},\mathbf{B}^{(t)}} \right| - \left| \sum_{k=1}^n s_k \beta_{\mathbf{B}^{(t)},\mathbf{B}^{(i)}} \right| \right) > |z_p| - |z_n| \right\}\notag
\\&~~~~~
P\left\{\left\|\boldsymbol{h}\right\|_2\left|\sum_{k=1}^ns_k\beta_{{\mathbf{B}^{(t)}},{\mathbf{B}^{(t)}}}\right|-|z_n|>0\right\}\notag
\\&~~~~~
+\operatorname*{P}\left\{\left(-\left\|\boldsymbol{h}\right\|_2\left|\sum_{k=1}^{n}s_{k}\beta_{{\mathbf{B}^{(t)}},{\mathbf{B}^{(t)}}}\right|+|z_{n}|>|z_{p}|+\left\|\boldsymbol{h}\right\|_2\left|\sum_{i=k}^ns_k\beta_{\mathbf{B}^{(t)},\mathbf{B}^{(i)}}\right|\right)\right\}\notag
\\&~~~~~
P\left\{\left\|\boldsymbol{h}\right\|_2\left|\sum_{k=1}^ns_k\beta_{{\mathbf{B}^{(t)}},{\mathbf{B}^{(t)}}}\right|-|z_n|<0\right\}\notag
\\&\ge
P\left\{ \left\| \boldsymbol{h} \right\|_2 \left( \left| \sum_{k=1}^n s_k \beta_{\mathbf{B}^{(t)},\mathbf{B}^{(t)}} \right| - \left| \sum_{k=1}^n s_k \beta_{\mathbf{B}^{(t)},\mathbf{B}^{(i)}} \right| \right) > |z_p| - |z_n| \right\}\notag
\\&~~~~~
P\left\{\left\|\boldsymbol{h}\right\|_2\left|\sum_{k=1}^ns_k\beta_{{\mathbf{B}^{(t)}},{\mathbf{B}^{(t)}}}\right|-|z_n|>0\right\},
\label{eq:final_match_ge_mismatched}
\end{align}
where 
\begin{align}
&|\left \langle {\Phi_{\mathbf{B}^{(t)}}} ,\mathbf{y}\right \rangle|
\ge 
\left\|\mathbf{h}\right\|_2\left|
\left\|\mathbf{h}\right\|  _2\left |   \sum_{k=1}^{n} s_k\cdot \beta_{{\mathbf{B}^{(t)}},{\mathbf{B}^{(t)}}}\right |-\left | z_n\right |
\right | \label{eq:scale_matched},
\\&|\left \langle {\Phi_{\mathbf{B}^{(i)}}} ,\mathbf{y}\right \rangle|
\le
\left\|\mathbf{h}\right\|_2\left(\left\|\mathbf{h}\right\|_2\left |   \sum_{k=1}^{n} s_k\cdot \beta_{{\mathbf{B}^{(t)}},{\mathbf{B}^{(i)}}}\right |+ \left|z_p\right|\right )
\label{eq:scale_mismatched}.
\end{align}

Since \(z_n\) follows a Gaussian distribution in \eqref{eq:z_np_distribution}, the term on the right-hand side of \eqref{eq:final_match_ge_mismatched} can be expressed as follows:
\begin{align}
&P\left\{\left\|\boldsymbol{h}\right\|_2\left|\sum_{k=1}^ns_k\beta_{{\mathbf{B}^{(t)}},{\mathbf{B}^{(t)}}}\right|-|z_n|>0\right\}\notag
\\&\ge 1-Q(\left\|\boldsymbol{h}\right\|_2\left|\sum_{k=1}^ns_k\beta_{{\mathbf{B}^{(t)}},{\mathbf{B}^{(t)}}}\right|)\notag
\\&\ge 1-\exp(-\frac{\left\|\boldsymbol{h}\right\|_2^2\left(\sum_{k=1}^ns_k\beta_{{\mathbf{B}^{(t)}},{\mathbf{B}^{(t)}}}\right)^2}{\sigma ^2} )\notag
\\& = 1-\exp(-\frac{\left\|\boldsymbol{h}\right\|_2^2\beta_{{\mathbf{B}^{(t)}},{\mathbf{B}^{(t)}}}^2\left(\sum_{k=1}^ns_k\right)^2}{\sigma ^2} ),\label{eq:right_final}
\end{align}
where \(Q(x)\le \exp(-\frac{x^2}{2})\) shown in \cite{SVC2018twc}.

Similarly, for the term of the right-hand side of \eqref{eq:final_match_ge_mismatched}, by relaxing the inequality within it, further deductions can be made as following
\begin{align}
&P\left\{ \left\| \boldsymbol{h} \right\|_2 \left( \left| \sum_{k=1}^n s_k \beta_{\mathbf{B}^{(t)},\mathbf{B}^{(t)}} \right| - \left| \sum_{k=1}^n s_k \beta_{\mathbf{B}^{(i)},\mathbf{B}^{(t)}} \right| \right) > |z_p| - |z_n| \right\}\notag
\\&
\ge P\left\{ \left\| \boldsymbol{h} \right\|_2 \left( \left| \sum_{k=1}^n s_k \beta_{\mathbf{B}^{(t)},\mathbf{B}^{(t)}} \right| - \left| \sum_{k=1}^n s_k \beta_{\mathbf{B}^{(i)},\mathbf{B}^{(t)}} \right| \right) > |z_p + z_n| \right\}\notag
\\&
\ge 
P\left\{ \left\| \boldsymbol{h} \right\|_2 \left( \left| \sum_{k=1}^n s_k (\beta_{\mathbf{B}^{(t)},\mathbf{B}^{(t)}}-\beta_{\mathbf{B}^{(i)},\mathbf{B}^{(t)}})\right| \right) > |z_p + z_n| \right\}\label{eq:left_term_in_final},
\end{align}
where 
\begin{align*}\left| \sum_{k=1}^n s_k \beta_{\mathbf{B}^{(t)},\mathbf{B}^{(t)}} \right| - \left| \sum_{k=1}^n s_k \beta_{\mathbf{B}^{(i)},\mathbf{B}^{(t)}} \right| \le \left| \sum_{k=1}^n s_k (\beta_{\mathbf{B}^{(t)},\mathbf{B}^{(t)}}-\beta_{\mathbf{B}^{(i)},\mathbf{B}^{(t)}})\right|.
\end{align*}

Since both \(z_p\) and \(z_n\) follow Gaussian distributions and are mutually independent, it follows that \(z_p+z_n\sim N(0,{\sigma^2})\). Therefore, we can further deduce \eqref{eq:left_term_in_final} that 
\begin{align}
&P\left\{ \left\| \boldsymbol{h} \right\|_2 \left( \left| \sum_{k=1}^n s_k (\beta_{\mathbf{B}^{(t)},\mathbf{B}^{(t)}}-\beta_{\mathbf{B}^{(i)},\mathbf{B}^{(t)}})\right| \right) > |z_p + z_n| \right\}\notag
\\&=1-Q(\left\| \boldsymbol{h} \right\|_2 \left( \left| \sum_{k=1}^n s_k (\beta_{\mathbf{B}^{(t)},\mathbf{B}^{(t)}}-\beta_{\mathbf{B}^{(i)},\mathbf{B}^{(t)}})\right| \right))\notag
\\&\ge 1-\exp(-\frac{\left\|\boldsymbol{h}\right\|_2^2\left(\sum_{k=1}^n s_k (\beta_{\mathbf{B}^{(t)},\mathbf{B}^{(t)}}-\beta_{\mathbf{B}^{(i)},\mathbf{B}^{(t)}})\right)^2}{2\sigma ^2} )\notag
\\&=1-\exp(-\frac{\left\|\boldsymbol{h}\right\|_2^2(\beta_{\mathbf{B}^{(t)},\mathbf{B}^{(t)}}-\beta_{\mathbf{B}^{(i)},\mathbf{B}^{(t)}})^2\left(\sum_{k=1}^n s_k \right)^2}{2\sigma ^2} )\label{eq:left_final}.
\end{align}

Hence, by substituting \eqref{eq:right_final} and \eqref{eq:left_final} into \eqref{eq:final_match_ge_mismatched}, the final expression can be derived as follows:
\begin{align}
&P(\mathbf{B}^{(t)}|\mathbf{h})
=\prod_{i\ne t}^{n} P(|\left \langle {\Phi_{\mathbf{B}^{(t)}}} ,\mathbf{y}\right \rangle| > |\left \langle {\Phi_{\mathbf{B}^{(i)}}} ,\mathbf{y}\right \rangle|)\notag
\\&=\left(1-\exp(-\frac{\left\|\boldsymbol{h}\right\|_2^2(\beta_{\mathbf{B}^{(t)},\mathbf{B}^{(t)}}-\beta_{\mathbf{B}^{(i)},\mathbf{B}^{(t)}})^2\left(\sum_{k=1}^n s_k \right)^2}{2\sigma ^2} )\right )^{n}\notag
\\&~~~~~~~
\left (1-\exp(-\frac{\left\|\boldsymbol{h}\right\|_2^2\beta_{{\mathbf{B}^{(t)}},{\mathbf{B}^{(t)}}}^2\left(\sum_{k=1}^ns_k\right)^2}{\sigma ^2} )\right )^{n}\notag
\\&\approx 
\left (1
-\exp(-\frac{\left\|\boldsymbol{h}\right\|_2^2(\beta_{\mathbf{B}^{(t)},\mathbf{B}^{(t)}}-\beta_{\mathbf{B}^{(i)},\mathbf{B}^{(t)}})^2\left(\sum_{k=1}^n s_k \right)^2}{2\sigma ^2} 
-\exp(-\frac{\left\|\boldsymbol{h}\right\|_2^2\beta_{{\mathbf{B}^{(t)}},{\mathbf{B}^{(t)}}}^2\left(\sum_{k=1}^ns_k\right)^2}{\sigma ^2} )\right )^{n}
\label{eq:final_p_b_h}
\end{align}

As \(\mathbf{h}=[h_1,~...~,h_n]\) is assumed to be a set of Gaussian random variables, which means \(h_{i}\sim \mathbb{C}N(\mu,\sigma^2)\), the inner product of the vector \(\mathbf{h}\) naturally follows a chi-squared distribution with \(n\) degrees of freedom, which means 
\begin{align*}
||\mathbf{h}||_2^2\sim \chi^2 (n)=\frac{x^{n-1}e^{-x}}{\Gamma(n)}, \Gamma(n)=(n-1)!
\end{align*}

Moreover, the expectation of the chi-squared distribution with respect to exponential functions always admits a closed-form analytical expression as shown in \cite{SVC2018twc}, which is 
\begin{align}
    \mathbb{E}_{\mathbf{h}}\left[
\exp(-\alpha x)\right] = 
\int _0^\infty \exp(-\alpha x )\frac{x^{m-1}\exp(-x)}{(m-1)!} dx
=
(1+\alpha)^{-m} \label{eq:e_exp}
\end{align}

Therefore, by combining \eqref{eq:final_p_b_h}, we can derive the following result:
\begin{align}
&P(\mathbf{B}^{(1)})=\mathbb{E}_{\mathbf{h}} [P(\mathbf{B}^{(t)}|\mathbf{h})]\notag
\\&\ge\mathbb{E}_{\mathbf{h}}\left [ \left (1-\exp(-\frac{\left\|\mathbf{h}\right\|_2^2(\beta_{\mathbf{B}^{(t)},\mathbf{B}^{(t)}}-\beta_{\mathbf{B}^{(t)},\mathbf{B}^{(i)}})^2\left(\sum_{k=1}^n s_k \right)^2}{2\sigma ^2} 
-\exp(-\frac{\left\|\mathbf{h}\right\|_2^2\beta_{{\mathbf{B}^{(t)}},{\mathbf{B}^{(t)}}}^2\left(\sum_{k=1}^ns_k\right)^2}{\sigma ^2} )\right )^{n}
 \right ] \notag
 \\&=
\left (1-\mathbb{E}_{\mathbf{h}}\left [ \exp(-\frac{\left\|\mathbf{h}\right\|_2^2(\beta_{\mathbf{B}^{(t)},\mathbf{B}^{(t)}}-\beta_{\mathbf{B}^{(t)},\mathbf{B}^{(i)}})^2\left(\sum_{k=1}^n s_k \right)^2}{2\sigma ^2}  \right ] 
-\mathbb{E}_{\mathbf{h}}\left [ \exp(-\frac{\left\|\mathbf{h}\right\|_2^2\beta_{{\mathbf{B}^{(t)}},{\mathbf{B}^{(t)}}}^2\left(\sum_{k=1}^ns_k\right)^2}{\sigma ^2} ) \right ] 
\right )^{n}\notag
\\&=
\left(
1-(
\frac{(\beta_{\mathbf{B}^{(t)},\mathbf{B}^{(t)}}-\beta_{\mathbf{B}^{(t)},\mathbf{B}^{(i)}})^2\left(\sum_{k=1}^n s_k \right)^2}{2\sigma ^2}
)^{-m}
-(\frac{(\beta_{\mathbf{B}^{(t)},\mathbf{B}^{(t)}})^2\left(\sum_{k=1}^n s_k \right)^2}{\sigma ^2})^{-m}
\right)^{n}\label{eq:final_p_b}
\end{align}

Therefore, by substituting \eqref{eq:final_p_b} into the initial success probability \eqref{eq:succ_prob}, we obtain the final result:
\begin{align*}
&P_{success} \approx \prod_{k=1}^K P(\mathbf{B}^{(k)})\\&\ge\left(1-(\frac{(\beta_{\mathbf{B}^{(t)},\mathbf{B}^{(t)}}-\beta_{\mathbf{B}^{(t)},\mathbf{B}^{(i)}})^2\left(\sum_{k=1}^n s_k \right)^2}{2\sigma ^2})^{-m}-(\frac{(\beta_{\mathbf{B}^{(t)},\mathbf{B}^{(t)}})^2\left(\sum_{k=1}^n s_k \right)^2}{\sigma ^2}
)^{-m}
\right)^{Kn}\end{align*}

In summary, the computational challenges being addressed in SMC differ significantly from traditional compressed sensing algorithms. This is primarily due to the excessively high matrix dimensions involved, as well as the specific structure of sparse values within the sparse vectors. Traditional compressed sensing algorithms typically deal with smaller matrices, allowing for computationally efficient column-wise matching. However, in the present context, the high matrix dimensions render the basic column-wise matching computationally prohibitive for large-scale data. Consequently, we employ block matching as a solution to this challenge. Block matching divides the large-scale data into smaller blocks for processing, thereby significantly reducing computational complexity and enhancing efficiency.

\bibliographystyle{IEEEtran}
\bibliography{refer}

\begin{thebibliography}{1}
\providecommand{\url}[1]{#1}
\csname url@samestyle\endcsname
\providecommand{\newblock}{\relax}
\providecommand{\bibinfo}[2]{#2}
\providecommand{\BIBentrySTDinterwordspacing}{\spaceskip=0pt\relax}
\providecommand{\BIBentryALTinterwordstretchfactor}{4}
\providecommand{\BIBentryALTinterwordspacing}{\spaceskip=\fontdimen2\font plus
\BIBentryALTinterwordstretchfactor\fontdimen3\font minus \fontdimen4\font\relax}
\providecommand{\BIBforeignlanguage}[2]{{%
\expandafter\ifx\csname l@#1\endcsname\relax
\typeout{** WARNING: IEEEtran.bst: No hyphenation pattern has been}%
\typeout{** loaded for the language `#1'. Using the pattern for}%
\typeout{** the default language instead.}%
\else
\language=\csname l@#1\endcsname
\fi
#2}}
\providecommand{\BIBdecl}{\relax}
\BIBdecl

\bibitem{SVC2018twc}
H.~Ji, S.~Park, and B.~Shim, ``Sparse vector coding for ultra reliable and low latency communications,'' \emph{IEEE Transactions on Wireless Communications}, vol.~17, no.~10, pp. 6693--6706, 2018.

\bibitem{pilotless2019LWC}
H.~Ji, W.~Kim, and B.~Shim, ``Pilot-less sparse vector coding for short packet transmission,'' \emph{IEEE Wireless Communications Letters}, vol.~8, no.~4, pp. 1036--1039, 2019.

\bibitem{SATO2021JIOT}
X.~Zhang, D.~Zhang, B.~Shim, G.~Han, D.~Zhang, and T.~Sato, ``Sparse superimposed coding for short-packet urllc,'' \emph{IEEE Internet of Things Journal}, vol.~9, no.~7, pp. 5275--5289, 2022.

\bibitem{SATO22022JCC}
X.~Zhang, H.~Chen, D.~Zhang, G.~Qin, B.~Davaasambuu, and T.~Sato, ``Uniquely decomposable constellation group-based sparse vector coding for short packet communications,'' \emph{China Communications}, vol.~20, no.~5, pp. 119--134, 2023.

\bibitem{Yang2023LCOMM}
L.~Yang and P.~Fan, ``Multiple-mode sparse superposed code with low block error rate for short packet urllc,'' \emph{IEEE Communications Letters}, vol.~28, no.~2, pp. 248--252, 2024.

\bibitem{optimalMat2023LWC}
------, ``Improved sparse vector code based on optimized spreading matrix for short-packet in urllc,'' \emph{IEEE Wireless Communications Letters}, vol.~12, no.~4, pp. 728--732, 2023.

\end{thebibliography}
\end{document}